\documentclass[prb,twocolumn,amssymb,showpacs]{revtex4}
\usepackage{graphicx}
\usepackage{amsmath}
\usepackage{txfonts}

\newlength{\figwidth}
 \divide\figwidth by 2
\newcommand{\tbti}{Tb$_{2}$Ti$_{2}$O$_7$}
\newcommand{\hoti}{Ho$_{2}$Ti$_{2}$O$_7$}
\newcommand{\dyti}{Dy$_{2}$Ti$_{2}$O$_7$}
\newcommand{\erti}{Er$_{2}$Ti$_{2}$O$_7$}
\newcommand{\ybti}{Yb$_{2}$Ti$_{2}$O$_7$}
\newcommand{\reti}{R$_{2}$Ti$_{2}$O$_7$}

\newcommand{\mub}{$\mu_B$}

\begin{document}

\title{Ising versus XY anisotropy in frustrated R$_{2}$Ti$_{2}$O$_7$ compounds as seen
 by polarized neutrons}
\author{H. Cao$^1$, A. Gukasov$^1$, I. Mirebeau$^1$, P. Bonville$^2$,
C. Decorse$^3$ and G. Dhalenne$^3$.
\\
}
\address{
$^1$Laboratoire L\'eon Brillouin, CEA-CNRS, CE-Saclay, 91191
Gif-sur-Yvette, France.}
\address {$^2$Service de Physique de l'Etat Condens\'e,
CEA-CNRS, CE-Saclay,  91191 Gif-Sur-Yvette, France.}
\address{$^3$Laboratoire de Physico-Chimie de l'Etat Solide, ICMMO, Universit\'e Paris-Sud, 91405 Orsay, France.}

\begin{abstract}
We studied the field induced magnetic order in R$_{2}$Ti$_{2}$O$_7$ pyrochlore
compounds with either uniaxial (R=Ho, Tb) or planar (R=Er, Yb) anisotropy, by
polarized neutron diffraction. The determination of the local susceptibility
tensor \{$\chi_\parallel,\chi_\perp$\} provides a universal description of the
field induced structures in the paramagnetic phase (2-270\,K), whatever the
field value (1-7\,T) and direction. Comparison of the thermal variations of
$\chi_\parallel$ and $\chi_\perp$ with calculations using the rare
earth crystal field shows that exchange and dipolar interactions must
be taken into account. We determine the molecular field tensor in each case
and show that it can be strongly anisotropic.

\end{abstract}

 \maketitle

The pyrochlore lattice of corner sharing tetrahedra offers the best model of
geometrical magnetic frustration in three dimensions.
In rare earth pyrochlore titanates \reti, frustration arises from the
subtle interplay of energy terms such as single ion anisotropy, exchange and
dipolar interactions. Depending on the balance between these factors, one may
observe spin ice (R=Ho, Dy)\cite{Harris98,Ramirez99,Bramwell01} or spin liquid
(R=Tb)\cite{Gardner99} behaviors, or complex magnetic orders stabilized by
first order transitions (R=Er, Yb, Gd)\cite{Champion03,Hodges02}.
A major role in the selection of the ground state is played by the single ion anisotropy that arises from the trigonal crystal
electric field (CEF) at the rare earth site. At low temperature, it
may result in strong axial or planar anisotropy, depending on the
wave function of the ground state and of the close CEF excited
states, if any.

It is now widely accepted that Ho, Dy and Tb titanates show Ising-like
behavior, the magnetic moments being constrained to align along the local
$<$111$>$ axes. The use of the Ising model for the
canonical spin ices \hoti\ and \dyti\ in a large temperature range is
justified by their extreme axial anisotropy and very high gap between the
CEF ground state and excited states \cite{Rosenkranz00,Jana00}.
In \tbti, where this energy gap is much smaller\cite{Gingras00,
Mirebeau07}, the Ising model holds only at low temperature;
above 200\,K, a Heisenberg-like behavior is recovered\cite{Kao03} since many
CEF levels become populated.
The spin liquid ground state in \tbti\ transforms into antiferromagnetic (AF) order
under applied pressure and/or magnetic field\cite{Mirebeau04,Rule06,Cao08}.
In zero field, none of these Ising titanates evidences long range
magnetic ordering (LRO) down to the lowest temperature.

The Er and Yb titanates,  where the easy plane for the
CEF ground state is the plane perpendicular to $<$111$>$
\cite{Champion03,Hodges01,Hodges02}, have been labelled XY-type
pyrochlores, although they do not present extreme planar anisotropy.
\erti\ was suggested to realize a model XY-type antiferromagnet, for
which theory predicts fluctuation-induced symmetry breaking, leading
to magnetic ordering. It indeed orders antiferromagnetically at
1.2\,K \cite{Champion03,Poole07}, but can be driven into a quantum
disordered state by application of a magnetic
field\cite{Ruff08,Clarty08}. In \ybti, a transition detected at 0.25\,K
\cite{Blote} was shown by $^{170}$Yb M\"ossbauer
spectroscopy and $\mu$SR\cite{Hodges02} to occur towards a short
range correlated state. This transition is first order and consists in a
sharp and strong decrease of the
spin fluctuation frequency. However, a time dependent ferromagnetic
order was also reported in \ybti \cite{Yasui03},
and the application of a moderate magnetic field was shown to drive it into a
stable LRO state \cite{Ross09}.

In the pyrochlore lattice, selection between Ising, Heisenberg or XY
models cannot be based, as usual, on the analysis of the macroscopic
properties of a single crystalline sample in a magnetic field
because of the
presence of four anisotropy axes, namely the $<$111$>$ axes.
Then only an average over the four local axes can be measured by
classical methods, and no direct information can be obtained about
the anisotropy of the local magnetic susceptibility of the R ion,
nor of the local exchange/dipolar tensor. There is thus a clear
motivation to get a direct access to these quantities and to
quantify the axial/planar anisotropy in the rare earth pyrochlores.

Here we show that the site susceptibility approach \cite{gukasov-brown}
developed for polarized neutron diffraction is an accurate tool for obtaining
the local susceptibility tensor {\boldmath $\chi$}. We present our results in
two Ising-type (\hoti, \tbti) and
two planar-type (\ybti, \erti) compounds in a large temperature
range (2\,K-270\,K). Knowledge of the {\boldmath $\chi$}-tensor allows one to
determine the moment values and
orientations in the field induced paramagnetic state, whatever the
field direction\cite{Cao08}. The measured thermal variations of the
components of {\boldmath $\chi$} are quantitatively compared with calculations
using CEF parameters deduced independently from inelastic neutron
scattering spectra. To obtain good agreement with the data, one needs to
introduce an anisotropic molecular field tensor, which takes into account
both the first neighbor exchange and the dipolar coupling.

Single  crystals of \hoti, \tbti, \erti\ and \ybti\ were grown 
by the floating-zone technique, using a mirror furnace. 
The crystals were characterized by zero field neutron diffraction at
100\,K and 5\,K. The nuclear structure factors were used to refine positional
parameters, occupancy factors, isotropic temperature
factors and extinction parameters, within 
the space group {\it Fd$\bar{3}$m}.

Neutron diffraction measurements were performed at the ORPH\'EE
reactor of the Laboratoire L\'eon Brillouin. Polarized neutron flipping
ratios were measured on the Super-6T2 spectrometer \cite{Super6T2} using
neutrons of incident wavelength $\lambda_n$=1.4\,\AA\ and polarization
$P_0$=0.98, and on the 5C1 spectrometer ($\lambda_n$=0.84\,\AA, $P_0$=0.91),
provided by a supermirror bender and a Heusler polarizer, respectively.
For each compound, 100-200
flipping ratios were measured, at selected temperatures in the range
2 - 270\,K, in a magnetic field of 1\,T applied  parallel to
the [110] direction. The minimum number of flipping ratios used in
the refinements with two parameters was 30. The programs CHILSQ \cite{ccsl}
were used for the least squares refinements of the  flipping ratios.

Assuming a linear response, the model assigns a susceptibility tensor
{\boldmath $\chi$}, of rank 3x3, to each crystallographically independent site
\cite{gukasov-brown,gukasov-rogl}. Its components $\chi_{ij}$ depend on
the site symmetry, and the magnetic moment $\bf{M}^d$ =
{\boldmath $\chi$} $\bf{H}$ induced on a R ion at the 16$d$ site in the unit
cell is not necessarily collinear with the field, since the off-diagonal
elements of {\boldmath $\chi$} can be non-zero. The tensor {\boldmath $\chi$}
has the same symmetry as the tensor {\boldmath $u$}
describing the thermal motion of atoms. The site susceptibility can
be conveniently visualized as a {\em magnetization ellipsoid}
constructed from the six $\chi_{ij}$ parameters in much the same way
as {\em thermal ellipsoids} are constructed from the $u_{ij}$
parameters. For a given ellipsoid, the radius vector gives the
magnitude and direction of the magnetic moment induced by a field of
1\,T rotating in space. For the  space group  {\it Fd$\bar{3}$m},
the symmetry constraints imply that {\boldmath $\chi$} has only
two independent matrix elements, $\chi_{11}$ and $\chi_{12}$.
At the 16$d$ site, the principal axes of the magnetization ellipsoid are
oriented along the four local $<$111$>$ axes. Their lengths are given by:
$\chi_\parallel=\chi_{11}+2\chi_{12}$ and
$\chi_\perp=\chi_{11}-\chi_{12}$.  These susceptibility components
 are determined for each temperature by refining the flipping ratios.

The thermal evolutions of $\chi_\parallel$ and $\chi_\perp$ are shown
in Figs.1-4 for the four compunds. In the insets are represented the susceptibility
ellipsoids at 5\,K (lower) and 250\,K (upper), whose axes have been
scaled by temperature to compensate for the Curie decrease, together
with the magnetic moments induced by the field applied along [110].
The four R sites of a tetrahedron are then separated into two classes: the
so-called $\alpha$-sites (the two on the left), where the angle between the
field and the ternary axis is 35.3$^\circ$, and the $\beta$-sites (the two on
the right), where the field is perpendicular to the ternary axis
\cite{Hiroi03}.
The data were interpreted using a single ion approach, where the trigonal
symmetry CEF and Zeeman Hamiltonian is written as:
\begin{equation}
{\cal H}_{CEF} = \sum_{n,m} B_n^m O_n^m - g_J\  \mu_B \ {\bf J.H}.
\label{eqcef}
\end{equation}
In this expression, the relevant coefficients are $B_2^0$, $B_4^0$, $B_4^3$,
$B_6^0$, $B_6^3$ and $B_6^6$, and the $O_n^m$ operators are expressed in
terms of spherical harmonics\cite{Hutchings,Wybourne}, which has the advantage
that the $B_n^m$ are roughly independent of the rare earth in
an isostructural series. In equation (\ref{eqcef}), $g_J$ is the Land\'e
factor of the rare earth. In order to take into account exchange and dipole-
dipole interactions, we introduced an anisotropic molecular field tensor
\{$\lambda_\parallel,\lambda_\perp$\} and performed a self-consistent
calculation, assuming a field component of 1\,T either parallel or
perpendicular to the ternary axis.

\begin{figure}
    \includegraphics* [width=\columnwidth] {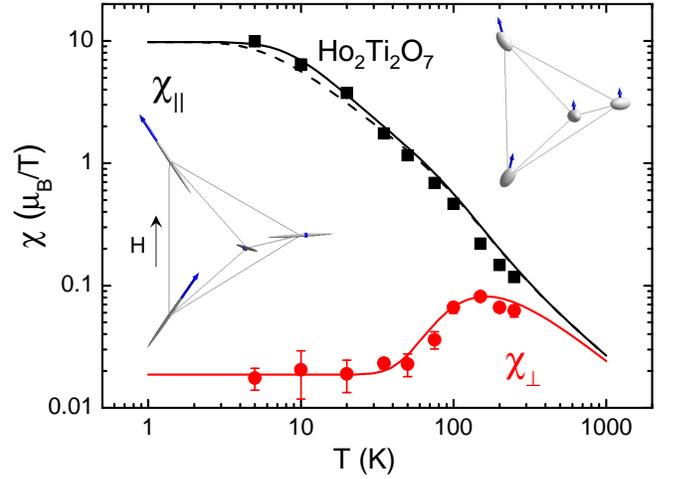}
\caption {(color on line). \hoti: Susceptibility components $\chi_\parallel$ and
$\chi_\perp$ versus $T$. The following CEF parameters (in K) were
used\cite{Rosenkranz00}: $B_2^0$ = 791, $B_4^0$ = 3188, $B_4^3$ = 971,
$B_6^0$ = 1007, $B_6^3$ = $-$725 and $B_6^6$ = 1178. Dashed lines:
CEF only calculation; solid lines: calculation including effective
exchange: $\lambda_\parallel$=0.05\,T/\mub, $\lambda_\perp$=0. For $\chi_\perp$, both lines superimpose.}
\label{chi_hoti}
\end{figure}

The data for \hoti\ are shown in Fig.\ref{chi_hoti}.
At low temperature, the ellipsoid elongations $\chi_\parallel$ increase so
markedly that they degenerate into
needles, while $\chi_\perp$ is very small, lying at the limit of
experimental precision. The induced magnetic structure is such that
the $\alpha$-site moments point towards the field direction, while
remaining along the local ternary axis, and the $\beta$-site moments
are zero. At high temperature, on the contrary, the anisotropy is
weaker and all the moments reorient close to the field direction.
These features, and in particular the peculiar shape of
$\chi_\perp(T)$, are explained by the CEF level scheme of Ho$^{3+}$
($J$=8, $g_J$=5/4 ) in \hoti: the ground state is very close to the
extremely anisotropic non-Kramers doublet $\vert J=8; J_z = \pm 8
\rangle$, which has a moment of 10\,\mub\ along the ternary axis and
$\chi_\perp = 0$, and the first excited state lies at about
250\,K
above the ground state\cite{Rosenkranz00,Jana00}. The lines in
Fig.\ref{chi_hoti} were calculated using the CEF parameters given in
Ref.\onlinecite{Rosenkranz00}.
We find that the CEF only curves (dashed
lines) yield a good agreement with the data. Due to the smallness of
$\chi_\perp(T)$, no reliable value for the molecular field constant
$\lambda_\perp$ can be obtained. For $\chi_\parallel$, introduction
of a small positive (ferromagnetic) value $\lambda_\parallel$=
0.05(1)\,T/\mub\ improves the agreement with the data (see
Fig.\ref{chi_hoti}). Since the saturated Ho magnetic moment along $<$111$>$ is
$\mu$ = 10\,\mub, this yields an effective field $H_\parallel =
\lambda_\parallel \mu =$0.5\,T. The exchange and dipolar
interactions in the ground spin-ice state of \hoti\ are well known
at low temperature\cite{bramharris01}, amounting to a positive
effective exchange constant $J_{eff} \simeq$ 1.8\,K, where the
dipolar contribution dominates. This is equivalent to a field
$H_{eff} \simeq$ 0.27\,T along the ternary axis. An exact comparison
with our result is however difficult because the magnetic structure
induced by the field in \hoti\ is quite different from the
zero-field moment arrangement in a spin-ice, hence modifying the
dipolar field at a Ho site.

\begin{figure}
\includegraphics* [width=\columnwidth] {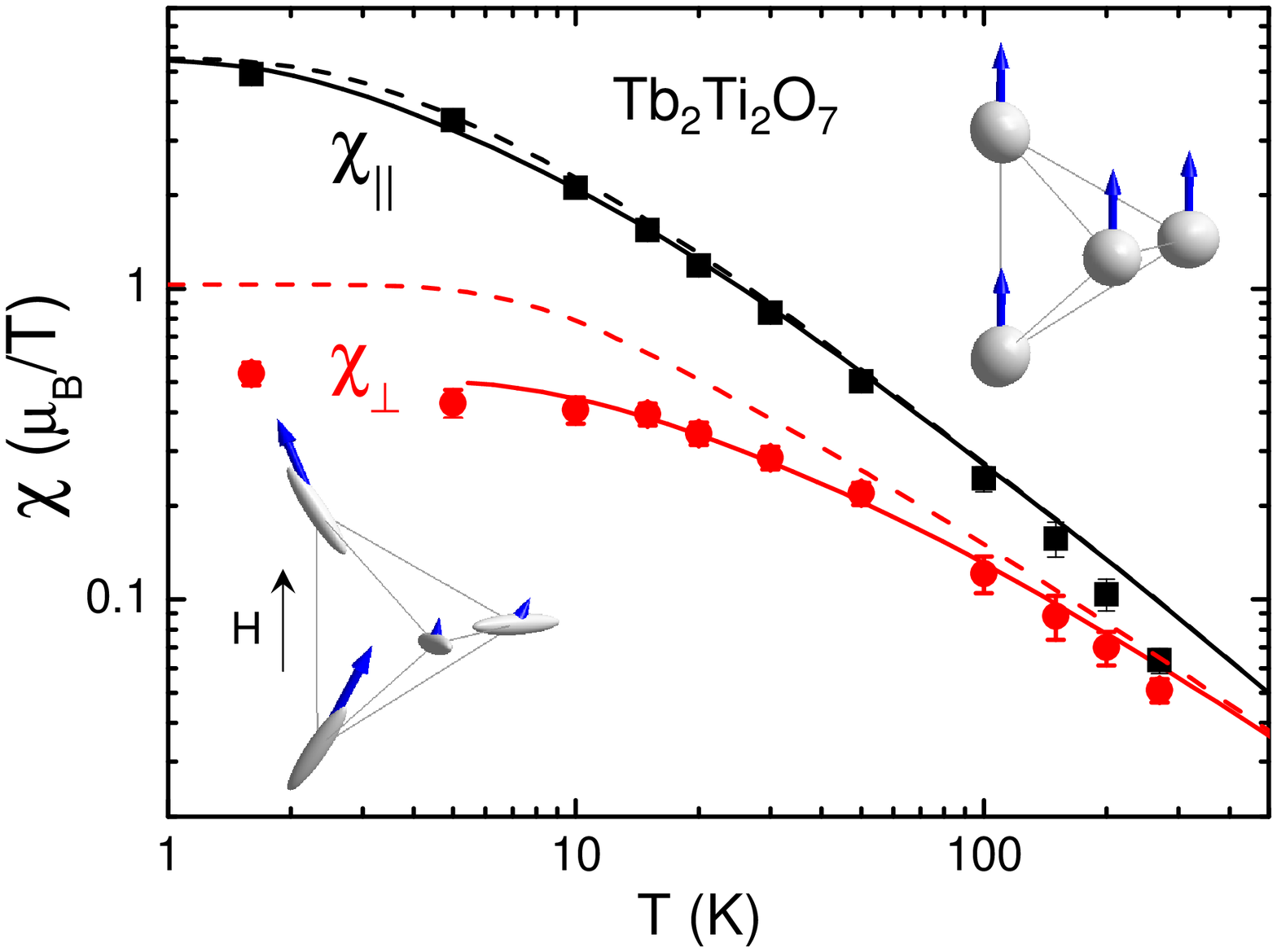}
\caption {(color on line). \tbti:
Susceptibility components $\chi_\parallel$ and $\chi_\perp$ versus
$T$. The following CEF parameters (in K) were used\cite{Mirebeau07}:
$B_2^0$ = 712, $B_4^0$ = 3400, $B_4^3$ = 1200, $B_6^0$ = 1130,
$B_6^3$ =$-$700 and $B_6^6$ = 1140. Dashed lines: CEF only
calculation; solid lines: calculation including effective exchange:
$\lambda_\parallel$=$-$0.05\,T/\mub, $\lambda_\perp$=$-$1\,T/\mub.}
\label{chi_tbti}
\end{figure}

The CEF anisotropy of Tb$^{3+}$ ($J$=6, $g_J$=3/2) in \tbti\ is
considerably smaller than that of Ho$^{3+}$ in \hoti\ (see
Fig.\ref{chi_tbti}). At room temperature, \tbti\ is practically
isotropic \cite{Kao03,Cao08,gukasovPNCMI} due to the
relatively large number of populated CEF levels
\cite{Gingras00,Mirebeau07}. Decreasing the temperature, it evolves
progressively into an Ising type with a ratio of ellipsoid axes of
the order of 10.
A CEF only calculation using the parameters from Ref.\onlinecite{Mirebeau07}
accounts practically for the $\chi_\parallel$ data, and lies somewhat above
the $\chi_\perp$ data. Introducing an anisotropic {\boldmath $\lambda$}-tensor
improves the agreement, especially for $\chi_\perp$, with the AF values:
$\lambda_\parallel$=$-$0.05(2)\,T/\mub\ and $\lambda_\perp$=$-$1.0(2)\,T/\mub.
We have checked that this {\boldmath $\lambda$}-tensor is coherent
with the isotropic effective value $\lambda=-$0.33\,T/\mub\ derived in
Ref.\onlinecite{Mirebeau07} from the analysis of the high temperature powder
susceptibility. For $\chi_\perp(T)$, the self-consistent calculation does not
converge below about 5\,K, which is the ``ordering'' temperature of the model;
this is why there are no calculated points for $\chi_\perp$ below 5\,K
in Fig.\ref{chi_tbti}. Both components of the {\boldmath $\lambda$}-tensor,
which corresponds to an effective (exchange + dipole) low temperature
interaction, are found to be of AF type, in line with the position of \tbti\
in the phase diagram of Ref.\onlinecite{hertog00}. Our own previous estimation
\cite{Mirebeau07} tentatively located \tbti\ in the spin ice side of this
phase diagram, but it was based on the high temperature isotropic determination of
$\lambda$ and could not take into account the anisotropy of the {\boldmath
$\lambda$}-tensor found in the present work.

\begin{figure}
\includegraphics* [width=\columnwidth] {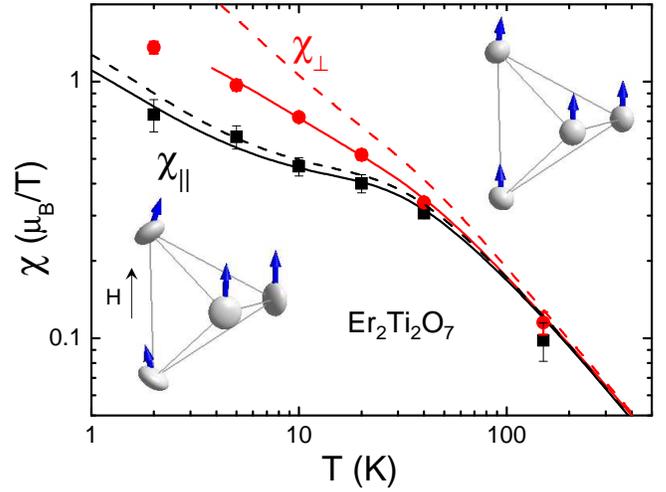}
\caption {(color on line). \erti: Susceptibility components $\chi_\parallel$ and $\chi_\perp$
versus $T$. The following CEF parameters (in K) were used (see text): $B_2^0$
=616, $B_4^0$ = 2850, $B_4^3$=795, $B_6^0$=858, $B_6^3$ = $-$493 and $B_6^6$
=980. Dashed lines:
CEF only calculation; solid lines: calculation including effective exchange:
$\lambda_\parallel$=$-$0.15\,T/\mub\ and $\lambda_\perp$=$-$0.45\,T/\mub.}
\label{chi_erti}
\end{figure}

As seen from Fig.\ref{chi_erti}, the anisotropy of Er$^{3+}$ ($J$=15/2, $g_J$
=6/5) in \erti\ is rather weak, and
a spherical shape for the magnetic ellipsoids is recovered above 100\,K. At
low temperature, $\chi_\perp > \chi_\parallel$, in agreement with the planar
magnetic ordering below 1.2\,K found in Ref.\onlinecite{Champion03}. The CEF
parameters for \erti, extrapolated from those in \hoti, were refined so as
to reproduce the energies of the two lowest CEF
excited doublets\cite{Champion03}, at 6.3 and 7.3\,meV, as well as the
transverse component of the $g$-tensor of the ground Kramers doublet, which
must match the in-plane saturated moment. The CEF parameters listed in the
caption of Fig.\ref{chi_erti} meet these requirements, with $g_\parallel$=2.6
and $g_\perp$ = 6.8 for the ground doublet, the latter value yielding
$m_{sat} = \frac{1}{2} g_\perp$ \mub = 3.4\,\mub, close to the measured value
3\,\mub. The CEF only calculation of
the susceptibilities shows a good first order agreement for
$\chi_\parallel(T)$, and introduction of
the effective exchange improves greatly the agreement for $\chi_\perp(T)$.
We find the {\boldmath $\lambda$}-tensor in \erti\ is also anisotropic and of
AF type: $\lambda_\perp$ = $-$0.45(5)\,T/\mub\ and $\lambda_\parallel$ = $-$
0.15(1)\,T/\mub. The in-plane effective exchange is thus stronger than its
component along the ternary axis, which reinforces the XY character determined
by the CEF anisotropy.
\begin{figure}
\includegraphics* [width=\columnwidth] {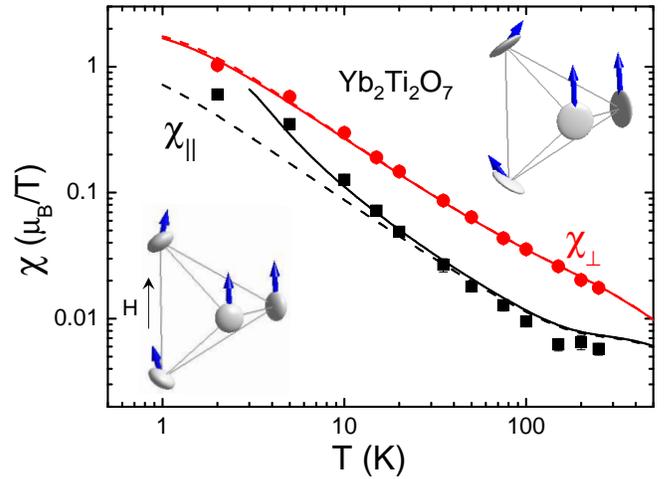}
\caption {(color on line). \ybti: Susceptibility components $\chi_\parallel$ and $\chi_\perp$
versus $T$. The following CEF parameters (in K) were used (see text): $B_2^0$
=750, $B_4^0$=2350, $B_4^3$=1200, $B_6^0$=700, $B_6^3$=$-$1000 and $B_6^6$=
700.
Dashed lines: CEF only calculations, solid lines: calculation including
effective exchange: $\lambda_\parallel$=2.5\,T/\mub, $\lambda_\perp$ =$-$
0.05\,T/\mub.}
\label{chi_ybti}
\end{figure}

Fig.\ref{chi_ybti} shows the temperature dependence of $\chi_\parallel$ and
$\chi_\perp$ for Yb$^{3+}$ ($J$=7/2, $g_J$=8/7) in the planar anisotropy
pyrochlore \ybti. Its magnetic
ellipsoids are disk-shaped, like in \erti, but a strong anisotropy persists up
to much higher temperature because the excited Kramers doublets lie much
higher in energy (700-1000\,K)\cite{Hodges01}. In order to calculate
$\chi_\parallel (T)$ and $\chi_\perp(T)$,
we used CEF parameters extrapolated from those in \hoti. The
chosen set of parameters must also reproduce the thermal
variation of the $4f$ quadrupolar moment $Q_{4f}(T) \propto \langle 3J_z^2 -
J(J+1) \rangle_T$ measured by the $\gamma-\gamma$ Perturbed Angular
Correlations technique \cite{Hodges01}. The $B_n^m$
coefficients listed in the caption of Fig.\ref{chi_ybti} meet these
requirements and yield a rather good agreement with the experimental data
for $\chi_\perp(T)$, implying $\lambda_\perp \simeq$ 0. For
$\chi_\parallel(T)$, a ferromagnetic value
$\lambda_\parallel$=2.5(5)\,T/\mub\ is needed to reproduce the data.
We checked that the powder susceptibility calculated using this {\boldmath
$\lambda$}-tensor agrees with experiment, i.e. that it leads to an almost
vanishing
paramagnetic Curie temperature below 20\,K \cite{Hodges01}. The present set
of parameters, which differs from the previous set
\cite{Hodges02} roughly by interchanging the $B_4^0$ and $B_6^0$ values,
is coherent with those in the \reti\ family. It yields a somewhat larger
$g_\parallel$ value for the ground doublet (2.25 $vs$ 1.80), which implies
that the Yb moment of 1.15\,\mub\ in the ``slow fluctuation'' phase below
0.25\,K \cite{Hodges02} probably lies along the $<$111$>$ axis.

\begin{table}[ht]
\centering
\begin{tabular} {|c|c|c|c|c|} \hline
        R           & Ho      & Tb         & Er         & Yb        \\ \hline \hline
$\lambda_\parallel$ & 0.05(5) & $-$0.05(2) & $-$0.15(1) & 2.5(5)
\\ \hline $\lambda_\perp$     & $--$    & $-$1.0(2)  & $-$0.45(5) &
0.00(5)  \\ \hline
\end{tabular}
\caption{Values of the two components of the molecular field tensor
(in T/\mub) in the \reti\ compounds.} \label{tablam}
\end{table}

In conclusion, polarized neutron diffraction allowed us to determine
the local susceptibility tensor in the \reti\ series, inaccessible by
macroscopic measurements in single crystals. Its temperature dependence in the
paramagnetic phase cannot be entirely accounted for by the sole
crystal field anisotropy: a molecular field tensor {\boldmath $\lambda$} must
be introduced, which encompasses exchange and dipolar interactions, and which
is found to be strongly anisotropic in most compounds.
The values of the {\boldmath $\lambda$} components are gathered in Table I.
Their sign and magnitude agree with the onset of planar
AF ordering (Er) and spin ice short range order (Ho). For Yb, we find
an extremely anisotropic ferromagnetic {\boldmath $\lambda$}-tensor along
$<$111$>$ with a moderate planar CEF anisotropy. By contrast, for Tb, which
has a strong CEF anisotropy along $<$111$>$, the {\boldmath $\lambda$}-tensor
is of AF type and characterized by a sizeable planar anisotropy. The
anisotropy of the {\boldmath $\lambda$}-tensor is likely to be due mainly to
the dipole-dipole coupling in the large moment
systems (Ho and Tb), but our results suggest that exchange itself is
anisotropic in the small moment systems (Er and Yb), where dipolar
interactions play a minor role. This could be an important ingredient for
theory in \tbti\ and
\ybti, where the ground state is not yet fully understood.

We thank Patrick Berthet for his support in the crystal synthesis, 
A. Cousson and B. Gillon for their help in the experiments. We also thank P. J. Brown for her advices in using CCSL and interesting
discussions. H. Cao acknowledges financial support from the Triangle de la
Physique.


\begin{thebibliography}{}

\bibitem{Harris98} M. J. Harris, S. T. Bramwell, P. C. W. Holdsworth, and
J. D. M. Champion,
Phys. Rev. Lett. \textbf{81}, 4496 (1998). 

\bibitem{Ramirez99} A. P. Ramirez {\em et al.} Nature \textbf{399}, 333
(1999).

\bibitem{Bramwell01} S. T. Bramwell and M. J. P. Gingras, Science
\textbf{294}, 1495 (2001).

\bibitem{Gardner99} J. S. Gardner {\em et al.},
 Phys. Rev. Lett. \textbf{82}, 1012 (1999).

\bibitem{Champion03} J. D. M. Champion {\em et al.} Phys. Rev. B {\bf 68},
020401(R) (2003).

\bibitem{Hodges02} J. A. Hodges {\em et al.}, Phys. Rev. Lett. \textbf{88},
077204 (2002).

\bibitem{Rosenkranz00} S. Rosenkranz, A. P. Ramirez, A. Hayashi, R. J. Cava,
R. Siddhartan and B. S. Shastry J. Appl. Phys. {\bf 87} 5914 (2000).

\bibitem{Jana00} Y. M. Jana and D. Ghosh Phys. Rev. B {\bf 61} 9657, (2000).



\bibitem{Gingras00} M. J. P. Gingras {\it et al} Phys. Rev. B {\bf 62}, 6496
(2000).

\bibitem{Mirebeau07} I. Mirebeau, P. Bonville and M. Hennion Phys. Rev. B
{\bf 76}, 184436 (2007).


\bibitem{Kao03} Y. J. Kao, M. Enjalran, A. Del Maestro,
H. R. Molavian, M. J. P. Gingras Phys. Rev. B. \textbf{68}, 172407 (2003).

\bibitem{Mirebeau04} I. Mirebeau, I. N. Goncharenko, G. Dhalenne and
A. Revcolevschi Phys. Rev. Lett. \textbf{93}, 187204 (2004).

\bibitem{Rule06} K. C. Rule {\em et al.}
Phys. Rev. Lett. {\bf 96}, 177201 (2006). 


\bibitem{Cao08} H. Cao, A. Gukasov, I. Mirebeau, P. Bonville and G. Dhalenne
Phys. Rev. Lett. {\bf 101}, 196402 (2008).

\bibitem{Hodges01} J. A. Hodges, P. Bonville, A. Forget, M. Rams,
K. Kr\'olas and G. Dhalenne J.Phys.: Condens.Matter \textbf{13}
9301, (2001).

\bibitem{Poole07} A. Poole, A. S. Wills and E. Leli\`evre-Berna, J.Phys.:
Condens. Matter {\bf 19}, 452201 (2007).

\bibitem{Ruff08} J. P. C. Ruff {\em et al.} Phys. Rev. Lett. {\bf 101}, 147205
(2008).

\bibitem{Clarty08} P. A. McClarty, S. H. Curnoe and M. J. P. Gingras
arXiv:0810.2483, cond-mat.str.-el. (2008)

\bibitem{Blote}
H. W. J. Bl\"ote, R. F. Wielinga, W. J. Huiskamp Physica \textbf{43}, 549
(1969).

\bibitem{Yasui03} Y. Yasui et al., J. Phys. Soc. Jpn. \textbf{72}, 3014 (2003).

\bibitem{Ross09} K. A. Ross {\em et al.}, arXiv:0902.0329, cond-mat.str.-el.
(2009)

\bibitem{gukasov-brown} A. Gukasov and P. J. Brown. J. Phys.: Condens. Matter
{\bf 14}, 8831, (2002).

\bibitem{Super6T2} A. Gukasov {\em et al.} 
Physica B \textbf{397}, 131–134 (2007).

\bibitem{ccsl} P. J. Brown and J. C. Matthewman, CCSL-RAL-93-009 (1993), and
http://www.ill.fr/dif/ccsl/html/ccsldoc.html

\bibitem{gukasov-rogl} A. Gukasov , P. Rogl, P. J. Brown, M. Mihalik and
A. Menovsky J. Phys.: Condens. Matter {\bf 14}, 8841 (2002).




\bibitem{Hiroi03}Z. Hiroi, K. Matsuhira and M. Ogata, J. Phys. Soc. Jpn.
\textbf{72}, 3045 (2003).
\bibitem{Hutchings}
M. T. Hutchings, Solid State Phys. \textbf{16}, 227 (1964).

\bibitem{Wybourne}
B. G. Wybourne, {\em Spectroscopic Properties of Rare Earths} (Interscience,
New York, 1965).

\bibitem{bramharris01}
S. T. Bramwell {\em et al}, Phys. Rev. Lett. \textbf{87}, 047205 (2001).

\bibitem{gukasovPNCMI} A. Gukasov, H. Cao, I. Mirebeau and P.
Bonville Physica B, Proceedings of PNCMI2008 (2009), in press.

\bibitem{hertog00} B. C. denHertog and M. J. P. Gingras, Phys. Rev. Lett.
\textbf{84}, 3430 (2000)

\end{thebibliography}
\end{document}